*Particle-mediated nucleation pathways are imprinted in the internal structure of calcium sulfate single crystals*

Tomasz M. Stawski[1]*, Helen M. Freeman[1,2], Alexander E.S. Van Driessche[3]**, Jörn Hövelmann[1], Rogier Besselink[1,3], Richard Wirth[1], and Liane G. Benning[1,4,5]

[1]German Research Centre for Geosciences, GFZ, Interface Geochemistry, Telegrafenberg, 14473, Potsdam, Germany;

[2]School of Chemical and Process Engineering, University of Leeds, Woodhouse Lane, LS2 9JT, Leeds, UK;

[3]UniversitéGrenoble Alpes, Université Savoie Mont Blanc, CNRS, IRD, IFSTTAR, ISTerre, 38000 Grenoble, France ;

[4]Department of Earth Sciences, Free University of Berlin, Malteserstr. 74-100 / Building A, 12249 , Berlin, Germany.

[5]School of Earth and Environment, University of Leeds, Woodhouse Lane, LS2 9JT, Leeds, UK.

Corresponding author(s):

*tomasz.stawski@gmail.com; www.researchgate.net/profile/Tomasz_Stawski

**alexander.van-driessche@univ-grenoble-alpes.fr

**Abstract**

Calcium sulfate minerals are found in nature as three hydrates: gypsum ($CaSO_4 \cdot 2H_2O$), bassanite ($CaSO_4 \cdot 0.5H_2O$), and anhydrite ($CaSO_4$). Due to their relevance in natural and industrial processes, the formation pathways of calcium sulfates from aqueous solution have been the subject of intensive research and there is a growing body of literature, suggesting that calcium sulfates form through a non-classical nanoparticle-mediated crystallisation process. We showed earlier (Stawski *et al.* 2016) that at the early stages in the precipitation reaction, calcium sulfate nano-crystals nucleate through the reorganization and coalescence of aggregates rather than through classical unit addition. Here, we used low-dose dark field (DF) transmission electron microscopy (TEM) and electron diffraction and document that these re-structuring processes do not continue until a final near-perfectly homogeneous single crystal is obtained. Instead we show that the growth process yields a final imperfect mesocrystal with an overall morphology resembling that of a single crystal, yet composed of smaller nano-domains. Our data reveal that organic-free calcium sulfate mesocrystals grown by a particle mediated-pathway may preserve in the final crystal structure a "memory" or "imprint" of their non-classical nucleation process, something that has been overlooked until now. Furthermore, the nano-scale misalignment of the structural sub-units within these crystals might propagate through the length-scales, which is potentially be expressed macroscopically as misaligned zones/domains in large single crystals. This is akin to observations in some of the giant crystals from the Naica Mine, Chihuahua, Mexico.

**Introduction**

Calcium sulfate minerals are abundant in natural and engineered environments and they exist in the form of three hydrates: gypsum ($CaSO_4 \cdot 2H_2O$), bassanite ($CaSO_4 \cdot 0.5H_2O$), and anhydrite ($CaSO_4$). Due to their relevance in natural and industrial processes, the formation pathways of these calcium sulphate phases from aqueous solution have been the subject of intensive research in the last few years[1]. Based on *in situ* and time-resolved small-angle X-ray scattering (SAXS) data, we recently documented how gypsum crystals form through the aggregation of sub-3 nm $CaSO_4$ precursor species to several-micron-large morphologies, and the consequent re-organisation, coalescence and growth of those precursor species within the aggregates[2]. Hence, the nucleation of gypsum is essentially a nanoparticle-mediated process. Importantly, faceted single crystals produced SAXS patterns, which at low-$q$ were characteristic for internally homogeneous large structures, yet at high-$q$ these patterns contained scattering features originating from nanosized sub-units[2]. Such scattering patterns can be fitted using a "brick-in-a-wall" surface fractal model, for which we developed a rigorous mathematical description[3]. This "brick-in-a-wall" model implies that sub-units constituting the bricks are clearly distinguishable from each other, leading to a single crystal composed of slightly misaligned crystallographic domains and hence expected to exhibit high mosaicity[4]. Hence, for nanoparticulate sub-units the "brick-in-a-wall" scattering model is in fact akin to the concept of mesocrystallinity[5–11].

Mesocrystals are single crystals in terms of their crystallographic properties and external forms, but they are internally composed of numerous crystalline nanoparticles or sub-domains of similar size and shape. These are arranged in a highly ordered, but spatially separated manner, yet the mesocrystals yield diffraction patterns typical for single crystals. Most commonly, mesocrystals are considered as composite assemblies of inorganic particles bound by organic species[11] (surfactants, macromolecules, small organic molecules etc.). However, mesostructured crystals can also form without the involvement of organic species. For example, this can occur through a process where stable pre-synthesised nanoparticles are "driven" to arrange themselves into larger crystals through physical fields such as magnetic, electric or hydration forces, so that they minimize their surface area, and thus lower their free energy[12–16]. Hence, in those cases growth proceeds through a non-classical particle-mediated bottom-up process, which can involve aggregation/re-orientation, and/or oriented attachment[6,17,18] of primary particles. Importantly, such organic-free mesocrystals can be necessary phases that are precursors to more internally continuous and stable single crystals, e.g., ferrihydrite transformation to goethite[17,19], or the formation of rutile from titania nanorods[20].

At present it is not clear what impact the particle-mediated crystallization pathway[7] has on the internal, and external structure of the final crystalline phase. We hypothesise that organic-free mesocrystal grown by a particle mediated-pathway may preserve a "memory" or "imprint" of this growth process in their final crystal structure. To test this hypothesis we built upon our previous work where we have shown that calcium sulphate formation from supersaturated aqueous solutions follows such a particle-mediated route. In the current study we characterised in detail the internal structure of different solid single crystals of the various calcium sulphate phases: synthetic gypsum and bassanite, as well as natural anhydrite (from the Naica Mine, Chihuahua, Mexico), using analytical transmission electron microscopy (TEM). We show compelling evidence for the mesostructured character of all these single crystals and consider the origin of their mesocrystallinity in the context of their growth mechanisms.

**Methods**

$CaSO_4 \cdot xH_2O$ single crystals in the form of gypsum (dihydrate, $2H_2O$, $x = 2$) and bassanite (hemihydrate, $0.5H_2O$, $x = 0.5$) were synthesized by reacting equimolar aqueous solutions of $CaCl_2 \cdot 2H_2O$ (pure, Sigma) and $Na_2SO_4$ (> 99%, Sigma), based on the following reaction:

$$Ca^{2+}_{(aq)} + 2Cl^-_{(aq)} + 2Na^+_{(aq)} + SO_4^{2-}_{(aq)} + xH_2O_{(l)} \rightarrow CaSO_4 \cdot xH_2O_{(s)} + 2Na^+_{(aq)} + 2Cl^-_{(aq)}$$

Gypsum formed at 21 °C from a solution with final [$CaSO_4$] concentration of 50 mmol/L, that was aged under stirring for 2 days. The so-formed gypsum crystals were dropcast onto holey carbon coated Cu TEM grids, dried in air and stored for further analyses. Bassanite was also synthesised from a 50 mmol/L [$CaSO_4$] solution, but with an increased salinity of 4.3 mol/L and at 80 °C, yet ageing was only for 8 hours. The high salinity/elevated temperature conditions promoted the reduced activity of water, which resulted in the direct precipitation of metastable hemihydrate, bassanite[21]. Following this synthesis step the solution containing the precipitated crystals were centrifuged at 3000 rpm and the supernatant was decanted. The remaining crystals were dried in air and then deposited onto a TEM grid without any dispersing medium prior to analyses.

These two synthetic $CaSO_4$ phases were complemented with large natural single crystals of gypsum and anhydrite ($x = 0$) that were obtained from the Niaca Mine in Chihauhua, Mexico[22]. The used specimens of gypsum and anhydrite were ~3 cm and ~1 cm in length respectively, and were both fragments chipped from larger single crystals (> 1 m in length for gypsum and >5 cm for anhydrite). To analyse these large natural single crystals we prepared ~15 µm x 4 µm thin foils (~100 nm) using the focused ion beam technique (FIB, FEI FIB200) following a standard procedure[23]. Unfortunately, in the case of the $CaSO_4$ dihydrate, gypsum, the FIB sample preparation step induced apparent damages to the hydrated crystals and the foils were visibly not stable during the further TEM analyses. Interestingly this was not the case for the anhydrous calcium sulfate, anhydrite, which was very stable under the vacuum conditions of sample preparation or TEM analyses; in addition, anhydrite was also not observably affected by the electron or ion beams at the operating conditions used. For comparison, we also prepared thin FIB foils of two natural olivine single crystals: (a) a Mg-rich olivine end-member, forsterite (formally $Mg_2SiO_4$,) and a mixed iron magnesium olivine (($Mg,Fe^{2+})_2SiO_4$) (SI: Fig. S1). Olivines are igneous minerals that are well-known to form naturally big and very high quality single crystals[24,25]. As such they constitute good examples of materials exhibiting very low mosaicity, as we discuss further.

For TEM imaging and selected-area electron diffraction (SAED), a Tecnai F20 XTWIN TEM was used at 200 kV, equipped with a field-emission gun electron source. Bright field (BF) and dark field (DF) images were acquired as energy-filtered images; for that purpose a 20 eV window was applied to the zero-loss peak. For DF TEM, the diffraction spots were selected by the objective aperture depending on the sample (the used diffraction spots are marked in Figs. 1-3 & SI: Fig. S1 accordingly). SAED patterns were collected using an aperture with an effective diameter of ~1 μm and the diffraction plates were developed in a high-dynamic range Ditabis Micron scanner. To correctly interpret any preferred orientation or texture-related effects in the TEM images, the objective astigmatism of the electron beam was corrected by ensuring the fast Fourier transform (FFT) was circular over the amorphous carbon film on which the synthetic single crystals or the FIB foils were deposited before collecting data from the samples. Selected images were initially analysed using ImageJ2[26] and any further processing was performed by means of bespoke scripts written in Python using NumPy, matplotlib and HyperSpy libraries[27–29].

**Results and Discussion**

*Synthetic gypsum and bassanite*

We employed a combination of high-resolution (HR), bright field (BF) and dark field (DF) TEM imaging and electron diffraction to explore the internal structure and crystallographic properties of the synthesized gypsum and bassanite single crystals from the meso- to nanoscales. Fig. 1A shows a low-magnification energy-filtered BF image of a synthetic gypsum single crystal that was aged in solution for 2 days at room temperature. The crystal is anisotropic in shape, has an elongated direction parallel to the *c*-axis and exhibits straight facets. The out-of-plane crystal thickness is highest in the centre (> 400 nm) as calculated by the log-ratio (relative) method[30] from the low-loss electron energy-loss spectra. This means that the bulk of the crystal in the field of view in Fig. 1A is not suitable for high resolution electron imaging, because it is too thick to obtain a sufficiently high signal-to-noise ratio from the CCD without significantly increasing the exposure time (which would inevitably cause beam damage to the material). However, the planes of the crystal facets do not intersect with each other at 90° (010, 120, -111 and 011 faces, see a schematic inset in Fig. 1A), but form a wedge, and therefore the external crystal perimeter is a much thinner region. The thickness contrast in the TEM image of a thin-edge region (parallel to the crystal long-axis) of the single crystal (Fig. 1B) gradually increases from right to left, which is caused by the aforementioned increase in thickness towards the central part of the crystal. The observed structure does not represent a typically expected homogeneous and continuous internal single crystal appearance. This is contrary to the data that we obtained from the single crystals of olivine that appear homogeneous (see SI: Fig. S1). In the synthetic gypsum single crystal in Fig. 1B, one can see that, within the field of view, the structure appears to be polycrystalline-like, where individual grains are distinguishable and exhibit a preferred orientation parallel to the *c*-axis of the crystal. This is confirmed by the fast Fourier transform (FFT, Fig. 1C), which shows an elliptically-shaped diffused low-angle scattering pattern localized around the centre of the image. Furthermore, in Fig. 1B only faint lattice fringes can be seen, which suggest that the individual grains are poorly crystalline, and/or that the orientation of some of the grains is mismatched. The corresponding FFT in Fig. 1C also shows one set of weak maxima belonging to the same *d*-spacing. This indicates that the grains exhibiting lattice fringes are crystallographically co-aligned. Overall this TEM analysis suggests that a gypsum single crystal is built of smaller nano-crystalline particles that are slightly misaligned with each other.

The SAED pattern of the crystal shown in Fig. 1A, confirms that the analysed crystal is a single gypsum crystal (Fig. 1D). The diffraction pattern contains only discrete diffraction peaks from a single crystallographic orientation. However, the recorded diffraction maxima exhibit very strong angle-dependent broadening effects exceeding >>1° at FWHMs. In single crystals, such effects, even when far smaller in magnitude, are typically attributed to a strong mosaicity[e.g., 31–33]. The mosaicity is a measure of the misalignment of crystallographic sub-domains building up a single crystal. In general terms, the division into sub-domains is a consequence of defects in the crystal lattice and does not necessarily mean that these domains physically constitute individual grain-like units. Nevertheless, the mosaicity in mesocrystals may be directly associated with the actual constituent nanosized sub-units. One can in fact identify the individual crystallographic domains in a mesocrystal, and hence evaluate their sizes, by performing DF TEM imaging with the diffracted beam corresponding to one of the diffraction maxima (Fig. 1D). Using this approach we show that a DF TEM image (Fig. 1E) of the single gypsum crystal represents its bright field counterpart from Fig. 1A. The high intensity (white) in the DF TEM image originates from the regions of interest, which are oriented in such a way that they fulfil the Bragg condition corresponding to a selected diffracted beam. Hence, for a homogeneous single crystal these regions should exhibit a high uniform intensity (as we confirm this with olivine; see SI: Fig. S1). Furthermore, for a crystal of gradually changing thickness, as in Fig. 1A, the intensity should gradually decrease with increasing crystal thickness. In order to be able to consider these two effects more clearly we enhanced the contrast in Fig. 1E using histogram equalization. This was performed locally with respect to the limited-size regions of the highest contrast (rather than the entire image). Additionally, the grey-scale intensity was remapped 1:1 to a fake-colour perceptually uniform 'inferno' scale[28] resulting in the enhanced image presented in Fig. 1F, where bright yellow/orange regions correspond to areas of high diffraction contrasts (bright yellow > orange). The intensity decreases gradually perpendicular to the perimeter of the crystal and its in-plane long axis. This is equivalent to the direction in which the thickness of the crystal increases (inset I, Fig. 1F) as evidenced in the BF TEM image (Fig. 1A). On the other hand, abrupt contrast variations parallel to the long central axis of the crystal, originate from the miss-alignment among the scattering domains (*i.e.*, mosaicity). Each individual bright spot represents a single scattering domain, which are discontinuously distributed and appear to be anisotropic in shape with their long axis orientated parallel to the long axis of the crystal. This is confirmed by the average of a 2D FFT series (inset II in Fig. 1F) calculated along the edges of the crystal (regions with the highest intensity). The compound FFT has an elliptical shape, rotated in such a way that its short axis is parallel to the long axis of the crystal, and thus corresponds to the long dimension of the anisotropic scattering domains in Fig. 1F. Thus, the FFT shown in inset II points to a preferred orientation of the scattering domains. The dimension of this scattering domains is ~10-20 nm (direction perpendicular to the long axis of the crystal). In fact the presence of such orientated anisotropic domains is also visible in the diffraction pattern in Fig. 1D, where the SAED contains characteristic streaks in [001] direction (marked with dashed green arrows). This again indicates the presences of very thin platelet-like or lamellae-like domains that are orientated in the direction parallel to the long axis of the crystal. On the whole, the results from the DF TEM corroborate those from BF TEM shown in Fig. 1B. The DF imaging is typically performed at low-magnification (i.e., low dose). This is highly beneficial, because TEM analysis (Fig. 1B) inherently increases the risk of beam damage in a highly hydrated sample such as gypsum[34]. We did not observe any obvious changes in the area imaged during our TEM measurements, but DF TEM further ensured that our observations did not contain any significant artifacts.

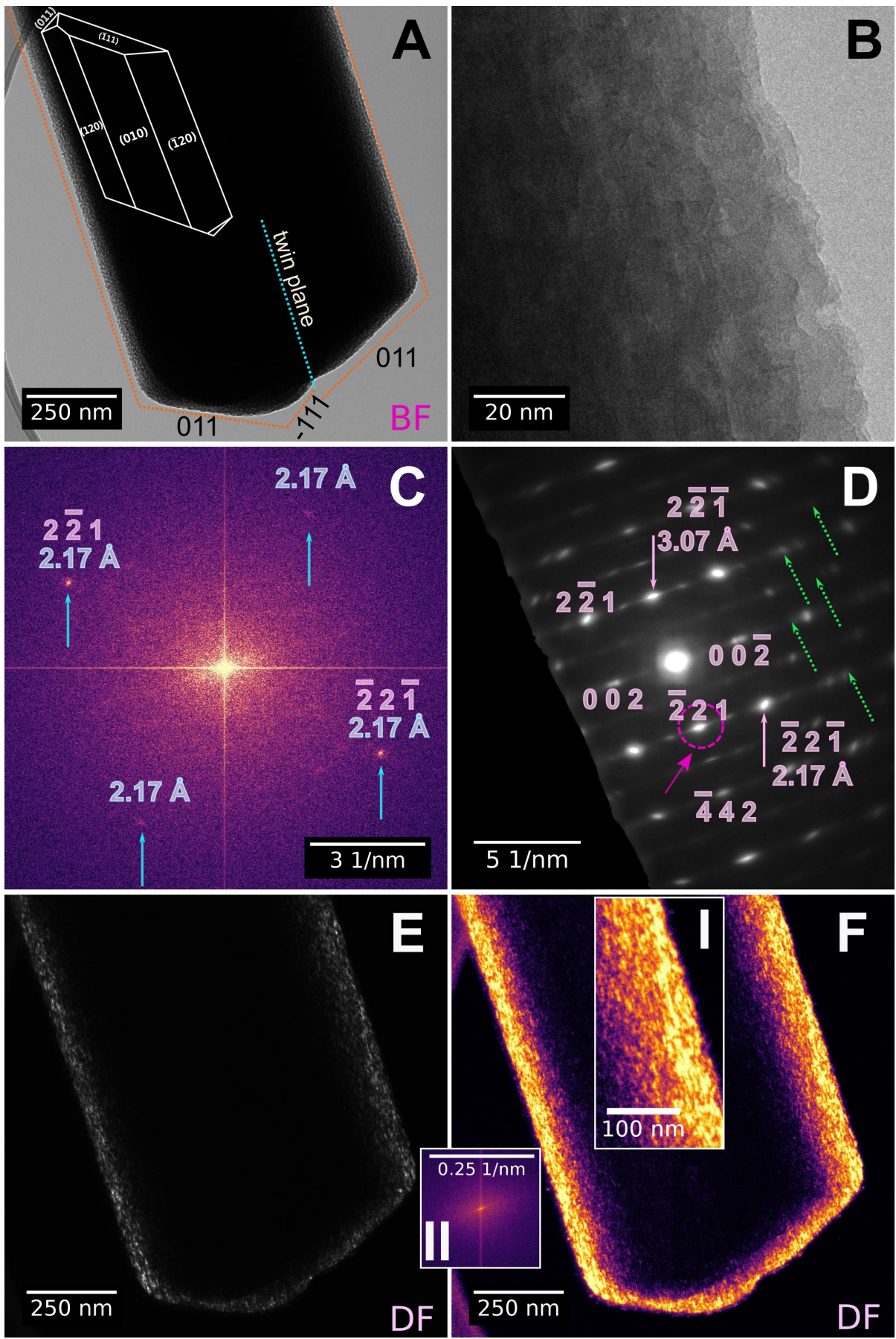

Fig. 1. TEM analysis of a representative gypsum crystal precipitated from a 50 mM CaSO$_4$ solution and equilibrated for 2 days with this solution at room temperature. A) BF TEM image; schematics of the planes of crystal facets characteristic for gypsum; flux: ~650 $e^-$Å$^{-2}$s$^{-1}$, estimated received fluence ~1 x 10$^{25}$ $e^-$/m$^2$; B) BF higher resolution TEM image from a thin region located on the right edge of the crystal; flux: ~9.4 x 10$^4$ $e^-$Å$^{-2}$s$^{-1}$, estimated received fluence ~1 x 10$^{26}$ $e^-$/m$^2$; C) fast Fourier transform (FFT) of image shown in (B); the indexed reflections indicated by the arrows and the corresponding *d*-spacings are characteristic for gypsum[35]; D) SAED pattern collected from the centre of the gypsum crystal shown in (A) with selected reflections and *d*-spacings labelled; the dashed circle marks the diffracted beam used for DF imaging; the dashed green arrows point to a characteristic streaking present in the diffraction pattern in [001] direction; E) unprocessed DF image corresponding to (A); F) DF image from (E) with enhanced contrast using a local histogram equalization technique and with the intensity remapped to an 'inferno' scale[28]; inset I – zoom-in of the selected region of (F) ; inset II – an average FFT calculated for the series taken along the left edge of the crystal.

To assess how the internal structure of single gypsum crystals compare to other calcium sulfate phases synthesized from solution we also analysed bassanite crystals (CaSO$_4$·0.5H$_2$O). Phase-pure bassanite can be directly made from solution by conducting the synthesis at low water activity[1,21,35–37]. We prepared bassanite samples following this strategy, where hemihydrate formed from a 50 mM CaSO$_4$ solution with very high salinity (4.3 M NaCl) at 80 °C (see Experimental and ref[21]). In Fig. 2A we present a BF image of a representative bassanite crystal that, similar to the gypsum crystal shown in Fig. 1, shows straight facets and a regular form. The analysis by SAED (Fig. 2B) confirmed this to be a single bassanite crystal and the individual diffraction spots again exhibited angle-dependent broadening. This diffraction pattern also exhibited streaking (dashed green arrows) in the [001] direction, which could be explained in terms of thin anisotropic subunits oriented parallel to the long axis of the crystal. The imaged bassanite crystal was significantly thinner than the gypsum crystal shown in Fig 1. The calculated thickness was only ~150 nm, and therefore the DF TEM image revealed a significantly higher level of detail from the interior of the crystal in addition to its perimeters (Fig. 2C). The corresponding contrast-enhanced image (Fig. 2D) shows that the bassanite single crystal was also composed of anisotropic nano-sized scattering domains aligned parallel to the long axis of the crystal (see FFT in inset I, Fig. 2D). These domains form locally parallel lines (inset II, Fig. 2D). Overall, the size and spatial arrangement of these domains are practically indistinguishable from those documented for the single gypsum crystal (Fig. 1).

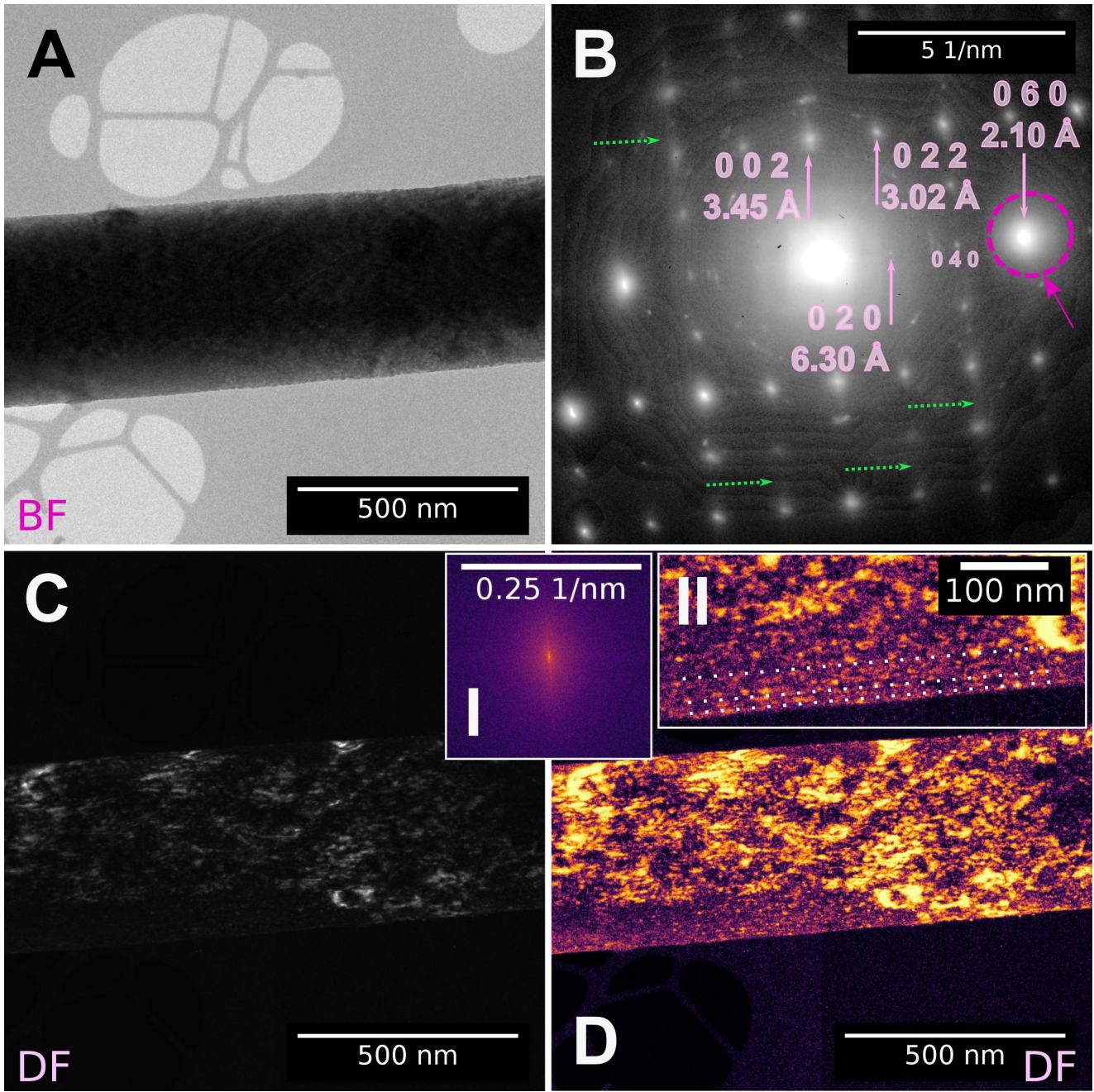

Fig. 2. TEM analysis of a representative bassanite crystal synthesised from CaSO$_4$ 50 mM and aged for 2 days in 4.3 M NaCl solution at 80 °C and aged for 8 hours. A) BF TEM image; flux: ~650 $e^-$Å$^{-2}$s$^{-1}$, estimated received fluence ~1 x 10$^{25}$ $e^-$/m$^2$; B) SAED taken from the centre of a bassanite crystal from (A); the indices and $d$-spacings of the selected diffraction

spots are provided; the dashed circle marks the diffracted beam used for dark field imaging; the dashed green arrows point to a characteristic streaking present in the diffraction pattern in [001] direction; C) unprocessed DF TEM image corresponding to (A); D) DF image with enhanced contrast and with an intensity remapped to an 'inferno' scale[28]; inset I – an average FFT calculated for the series taken along the entire long axis of the crystal; inset II – blow-up of the selected region of the main image. Individual scattering domains form parallel lines some of which are highlighted by dotted lines for ease of viewing.

*Natural calcium sulfate phases*

Under natural conditions calcium sulfate phases, in particular gypsum and anhydrite, are known to grow into single crystals that can easily reach many centimetres or even metres in size[22,38–40]. The question arises if such big natural crystals also grow and develop structures resembling those presented above for the synthetic phases. To test this we analysed natural single crystals of anhydrite from which we cut out FIB foils and analysed them by TEM (see Experimental: gypsum FIB foils were unstable under TEM/FIB). Fig. 3A shows a BF image of such an anhydrite thin foil. The observed contrast variations at the length-scale of 10-20 nm, form a regular pattern, which we attribute to the defect structure of the material. The SAED in the inset confirms that we are dealing with a single crystal of anhydrite, similar to the patterns obtained for the synthetic calcium sulfate crystals of gypsum and bassanite (Figs. 1&2). The elliptical shape of the diffraction spots in the SAED (Fig 3A, inset) points to a significant mosaicity. DF imaging (Fig. 3B) highlights ~10-20 nm sized discontinuous diffraction domains within this crystal, which coincide with the microstructural pattern observed in BF TEM (Fig. 3A). This is in stark contrast with the FIB foils from the single crystals of olivine (SI: Fig. S1), where the crystallographic domains are continuous throughout the micron length-scales of the region of interest. Similar to the synthetic gypsum and bassanite (Figs. 1&2), the nano-domains in the natural anhydrite are aligned along a single direction as confirmed by the anisotropic shape of the FFT obtained from the DF TEM image (inset in Fig. 3B). The high stability of the anhydrite thin-foil allowed us to perform high-resolution imaging (Fig. 3C), which confirmed the overall single-crystalline nature of the anhydrite sample (see FFT in the inset in Fig. 3C). However, the HR TEM image also contains clear areas that are less ordered then the surrounding crystalline areas. We highlighted those by performing FFT filtering of the image in Fig. 3C and then performing the inverse FFT reconstruction and applying a fake colour map. The resulting Fig. 3D highlights the nanocrystalline domains (10-20 nm in size) separated from each other by disordered areas of several nanometres in width.

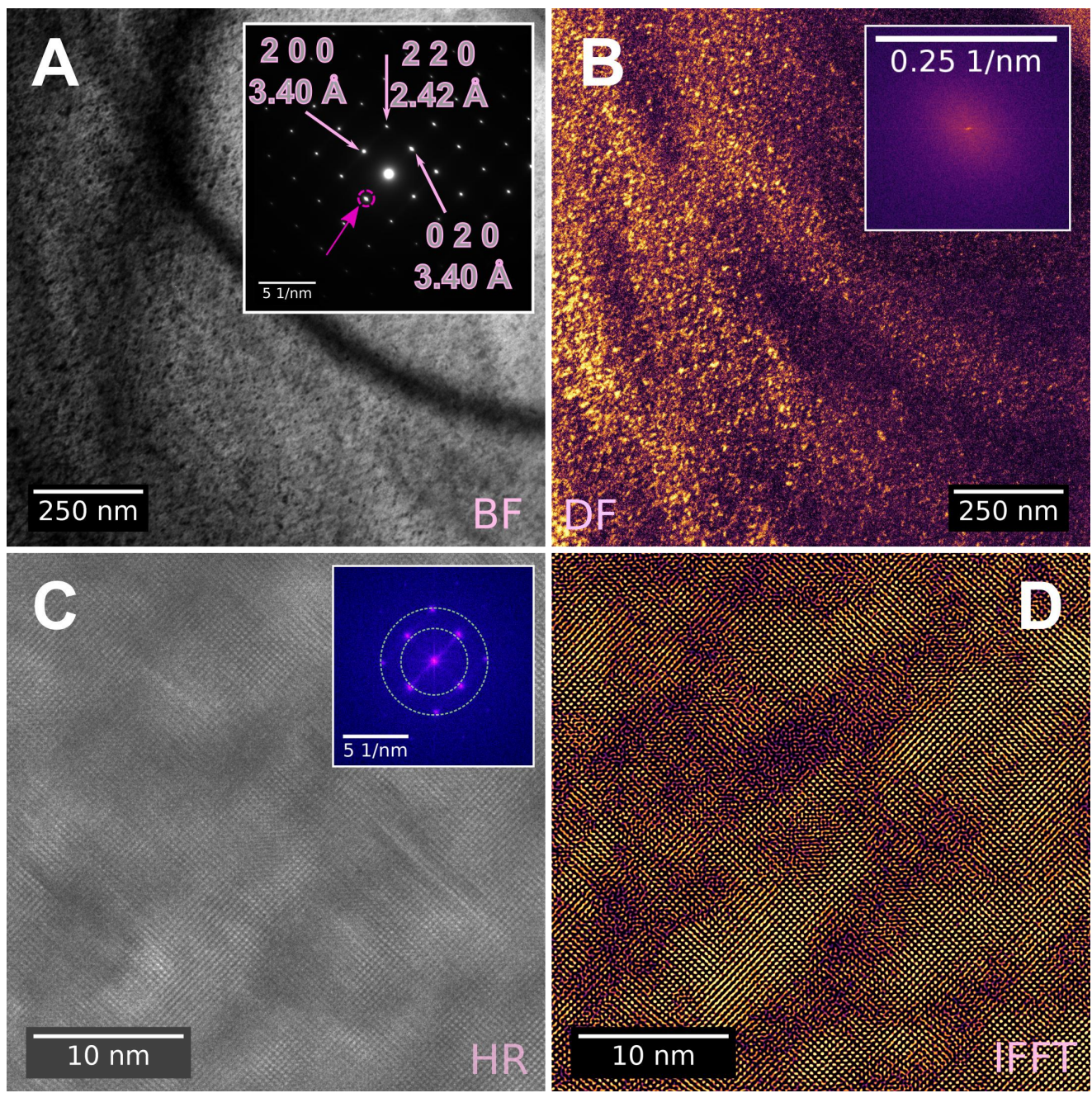

Fig. 3. TEM analysis of a FIB-foil cut from a natural anhydrite single crystal (see Fig. S2). A) BF-TEM image; flux: ~650 $e^-$Å$^{-2}$s$^{-1}$, estimated received fluence: 2 x 10$^{25}$ $e^-$/m$^2$; the inset shows an SAED pattern taken from the field of view; the dashed circle marks the diffracted beam used for DF imaging; the indices and *d*-spacings of the selected diffraction spots are provided; B) DF TEM image corresponding to (A) with enhanced contrast and with the intensity remapped to an 'inferno' scale[28]; the inset shows an FFT calculated for the image in (B); C) the HR TEM image of a small section of the foil; flux: ~8 x 10$^5$ $e^-$Å$^{-2}$s$^{-1}$, estimated received fluence ~1 x 10$^{27}$ $e^-$/m$^2$; the inset shows the FFT of the HR TEM image and indicates that the lattice fringes in (C) originate from a single orientation of a crystal; the green dashed circles mark the inner and the outer diameters of the filter ring used to obtain (D); D) the FFT-filtered image from (C) (inverse FFT), with an 'inferno' colour map applied.

**Mechanisms, Implications and Outlook**

Our microscopic analysis of both lab-grown and natural single crystals of the various calcium sulfate phases shows that they all share remarkably similar nano/microstructures. The considered single crystals of gypsum, bassanite and anhydrite are all composed of slightly misaligned anisotropic crystallographic domains, which are ~10-20 nm in size. Hence, we classify these crystals as mesocrystals. Following the current consensus discussed above, we point out that this classification is solely based on their final structure, which essentially does not provide on its own a sufficient evidence for a non-classical, particle-mediated, growth mechanism[41,42]. However, in our previous work based on scattering experiments and theory[1–3,43], we showed that the formation of calcium sulfate phases occurs through the coalescence and growth of primary particles within surface fractal aggregates ("brick-in-a-wall")[3]. We postulated that this initial step involved a framework structure as a plausible common precursor to gypsum, bassanite and anhydrite[1,2]. Our scattering data showed that the primary particles in these framework structures are nano-sized Ca-SO$_4$ clusters (<3 nm in length). In the initial growth stages the clusters aggregate rapidly to several-micron-sized morphologies to form surface fractal aggregates. During the actual crystallization process these aggregated clusters increase in dimension and polydispersity, undergo coalescence and form larger structural nanoparticulate sub-units within the aggregates which transform into the final crystals. Important is the fact that after the crystallization onset these structural sub-units are still distinguishable, compellingly indicating that misalignments and voids exist between the sub-units, similar to what we documented in the natural anhydrite sample (Fig. 3D). This comparison is exemplified in Fig. 4 where we show a SAXS pattern 4 hours after the onset of gypsum crystal nucleation and growth[1,2,43]. The 2D SAXS patterns were anisotropic (Fig. 4A), i.e., stronger scattering at higher angles was observed in the direction almost parallel to the *y* axis of the 2D detector plane (vertical direction), and thus normal to the *x* axis (horizontal direction). This anisotropy is further highlighted by the reduction in the direction-dependent 1D scattering patterns (Fig. 4B). The presented SAXS measurements were taken from a solution flown through the horizontally mounted capillary[2]. For the used experimental conditions gypsum forms high-aspect-ratio elongated crystals, i.e., needles[35], whose long axis became aligned with the flow. Such an anisotropic scattering pattern is expected if within the accessible *q*-range there are orientation-dependent internal variations in the microstructure of the needle-shaped crystalline particles with respect to their long-axes. Thus, these aged gypsum crystals are composed of smaller structural features (e.g., nanoparticles), which are oriented with respect to each other. The larger dimension of these anisotropic nano-domains are oriented parallel to the flow and the long-axis of the crystals containing these particles, and *vice versa* (inset in Fig. 4B). The intensity of both direction-dependent 1D scattering patterns, i.e., parallel and perpendicular to the flow direction, scale with $\sim q^{-4}$ for $q_0 < \sim 0.5$ nm$^{-1}$ (Fig. 4B), originating from the internal and external interfaces of the gypsum crystals. For $q_0 > \sim 0.5$ nm$^{-1}$ the scattering intensity of both patterns scales with $q^{>-4}$, originating from structural features with a characteristic size of $\sim 2\pi/q_0 = \sim 13$ nm (equivalent to the radius of gyration). Therefore, we defined these crystals as "brick-in-a-wall" surface fractal aggregates[3]. At long length-scales (low-*q*) these objects appear to be homogeneous (single crystals) with their scattering patterns dominated by the interface between the them and the surrounding solution (i.e., $I(q) \propto q^{-3 > a \geq -4}$). In contrast, at short length-scales (high-*q*), the scattering patterns represent form factors of the building units of these crystals. These are very densely packed with respect to each other (in contrast to classical mass fractals), yet still clearly distinguishable.

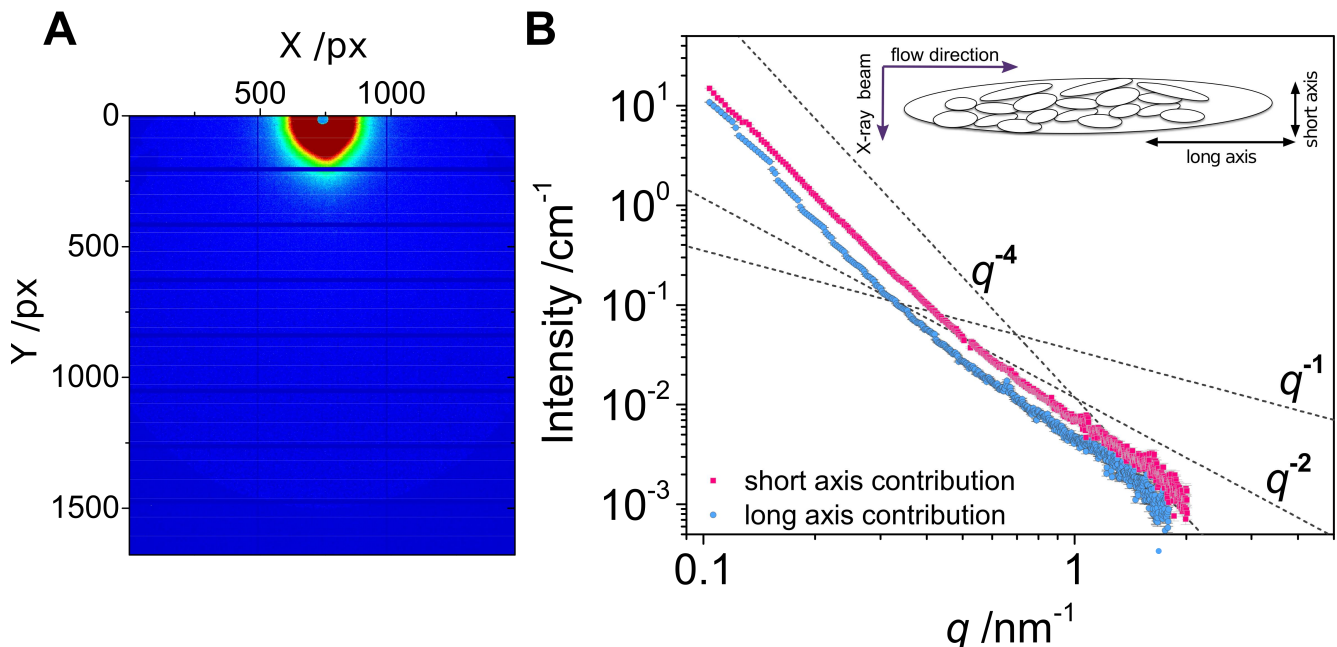

Fig. 4., *In situ* SAXS data acquired from laboratory synthesised gypsum crystals measured using a flow-cell (adapted from [2]). A) 2D SAXS patterns from a 50 mM $CaSO_4$ solution with gypsum crystals equilibrated at 12 °C for 4 hours after the onset of precipitation; intensity scale color-coding: red – high, blue – low; B) 1D angle-dependent SAXS curves from (A) obtained by averaging pixels at similar $q$ and limited to ca. +/- 3° angle off the direction indicated by the chosen azimuthal angle (the equatorial and meridional directions of the elliptical 2D pattern). The change in the $I(q)$ dependence of the scattering exponent in different parts of the 1D patterns are also shown to emphasize the differences in the high-q part of the data (dashed lines). Inset: schematic representation of the morphology of flow-oriented particles.

Our data show that the re-structuring and coalescence processes by no means continue until a near-perfectly homogeneous single crystal is obtained, and instead they come to a stop or at least significantly slow down. This growth behaviour can be rationalized if we consider that during the early stages of precipitation all calcium sulfate crystals appear to grow through the reorganization and coalescence of the primary species within the aggregates rather than through unit addition. In order to obtain well-ordered anhydrous cores of Ca-SO4 surrounded by $H_2O$ layers (as found in gypsum), $H_2O$ channels (bassanite), or fully dehydrated crystals (anhydrite), the nanoparticle aggregates must radically transform from a local less-ordered structure to a more ordered crystal. Hence, any mass transport processes inside such aggregates must be subject to slow diffusional processes compared to ion transport through the bulk aqueous solution. It is likely that the decreasing diffusion rate upon crystallisation inside the aggregates limits the domains' sizes to the observed 10 - 20 nm. Such a process thus yields a final imperfect mesocrystal, composed of smaller domains rather than a perfectly continuous single crystal. This early-stage crystallization however, does not exclude growth by ion-by-ion addition[39], a process that will still dominate at the latter stages of the crystallisation process (and thus potentially might yield more structurally homogeneous single crystals). During this secondary ion-by-ion growth the original non-classical mesostructured single crystals would constitute imperfect, i.e. slightly misaligned, seeds. Hence, the nano-scale misalignment of the structural sub-units observed in the above discussed crystals might propagate through the length-scales, and this could be expressed macroscopically as spherulites (formed at low supersaturations) and multiple twins (Figs. 5A&B), or as misaligned zones/domains in large single crystals, as one can clearly see in the giant crystals from Naica grown at extreme low levels of supersaturation[38] (Fig. 5C).

In conclusion, the data presented above provide compelling evidence that in calcium sulfate phases a particle-mediated nucleation pathway is essentially "frozen in", i.e., the final crystals are retaining the initial nanoparticle aggregate structure. This finding is paramount to explain the patterns observed of natural calcium sulfate formation, but is also essential to improve our control over the crystallisation of calcium sulfate phases, an industrially relevant material (e.g., *plaster of Paris*, scalants).

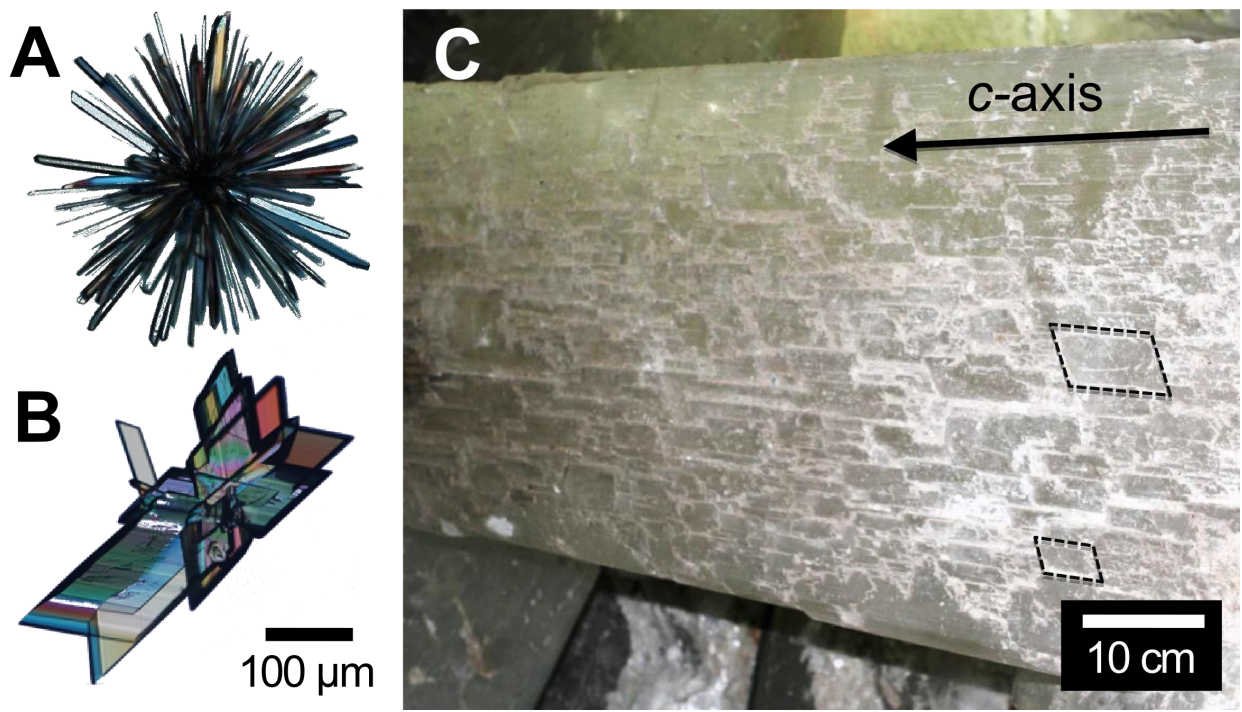

Fig. 5. Typical (A) spherulite and (B) multiple twin morphologies observed for lab grown gypsum crystals at the same magnification; (C) A metre-sized single crystal of gypsum from the Naica Mine, Chihuahua, Mexico. Smaller co-aligned domains (a few are delineated with dotted black parallels) are clearly visible in the bulk of the crystal. An arrow indicates the *c*-axis of the crystal.

**Supporting Information summary**

- Figure S1. TEM analysis of a FIB thin-foil cut from a natural forsterite (nominally $Mg_2SiO_4$) and generic olivine $((Mg,Fe^{2+})_2SiO_4)$ single crystals.

**Acknowledgments**

This research was made possible by a Marie Curie grant from the European Commission: the NanoSiAl Individual Fellowship, Project No. 703015 to TMS. We also acknowledge the financial support of the Helmholtz Recruiting Initiative grant No. I-044-16-01 to LGB. AESVD acknowledges financial supported by the French national programme EC2CO - Biohefect, SULFCYCLE.

**References**

(1) Van Driessche, A. E. S.; Stawski, T. M.; Benning, L. G.; Kellermeier, M. Calcium Sulfate Precipitation Throughout Its Phase Diagram. In *New Perspectives on Mineral Nucleation and Growth*; Springer International Publishing: Cham, 2017; pp 227–256.

(2) Stawski, T. M.; van Driessche, A. E. S.; Ossorio, M.; Diego Rodriguez-Blanco, J.; Besselink, R.; Benning, L. G. Formation of Calcium Sulfate through the Aggregation of Sub-3 Nanometre Primary Species. *Nat. Commun.* **2016**, *7*, 11177.

(3) Besselink, R.; Stawski, T. M.; Van Driessche, A. E. S. S.; Benning, L. G. Not Just Fractal Surfaces, but Surface Fractal Aggregates: Derivation of the Expression for the Structure Factor and Its Applications. *J. Chem. Phys.* **2016**, *145* (21), 211908.

(4) Darwin, C. G. The Reflexion of X-Rays from Imperfect Crystals. *London, Edinburgh, Dublin Philos. Mag. J. Sci.* **1922**, *43* (257), 800–829.

(5) Cölfen, H.; Yu, S.-H. Biomimetic Mineralization/Synthesis of Mesoscale Order in Hybrid Inorganic–Organic Materials via Nanoparticle Self-Assembly. *MRS Bull.* **2005**, *30* (10), 727–735.

(6) Niederberger, M.; Cölfen, H. Oriented Attachment and Mesocrystals: Non-Classical Crystallization Mechanisms Based on Nanoparticle Assembly. *Phys. Chem. Chem. Phys.* **2006**, *8* (28), 3271–3287.

(7) De Yoreo, J. J.; Gilbert, P. U. P. A.; Sommerdijk, N. A. J. M.; Penn, R. L.; Whitelam, S.; Joester,

D.; Zhang, H.; Rimer, J. D.; Navrotsky, A.; Banfield, J. F.; et al. Crystallization by Particle Attachment in Synthetic, Biogenic, and Geologic Environments. *Science* **2015**, *349* (6247), aaa6760.

(8)  Zhou, L.; O'Brien, P. Mesocrystals: A New Class of Solid Materials. *Small* **2008**, *4* (10), 1566–1574.

(9)  Sturm (née Rosseeva), E. V.; Cölfen, H. Mesocrystals: Structural and Morphogenetic Aspects. *Chem. Soc. Rev.* **2016**, *45* (21), 5821–5833.

(10) Coelfen, H.; Antonietti, M. *Mesocrystals and Nonclassical Crystallization: New Self-Assembled Structures*; Wiley, 2008.

(11) Rao, A.; Cölfen, H. Mineralization Schemes in the Living World: Mesocrystals. In *New Perspectives on Mineral Nucleation and Growth*; Springer International Publishing: Cham, 2017; pp 155–183.

(12) Agthe, M.; Wetterskog, E.; Mouzon, J.; Salazar-Alvarez, G.; Bergström, L. Dynamic Growth Modes of Ordered Arrays and Mesocrystals during Drop-Casting of Iron Oxide Nanocubes. *CrystEngComm* **2014**, *16* (8), 1443–1450.

(13) Yang, P.; Kim, F. Langmuir - Blodgett Assembly of One-Dimensional Nanostructures. *ChemPhysChem*. Wiley-Blackwell June 17, 2002, pp 503–506.

(14) Li, L.; Yang, Y.; Ding, J.; Xue, J. Synthesis of Magnetite Nanooctahedra and Their Magnetic Field-Induced Two-/Three-Dimensional Superstructure. *Chem. Mater.* **2010**, *22* (10), 3183–3191.

(15) Bergström, L.; Sturm née Rosseeva, E. V.; Salazar-Alvarez, G.; Cölfen, H. Mesocrystals in Biominerals and Colloidal Arrays. *Acc. Chem. Res.* **2015**, *48* (5), 1391–1402.

(16) Ma, M. G.; Cölfen, H. Mesocrystals - Applications and Potential. *Current Opinion in Colloid and Interface Science*. Elsevier April 1, 2014, pp 56–65.

(17) Penn, R. L.; Li, D.; Soltis, J. A. A Perspective on the Particle-Based Crystal Growth of Ferric Oxides, Oxyhydroxides, and Hydrous Oxides. In *New Perspectives on Mineral Nucleation and Growth*; Springer International Publishing: Cham, 2017; pp 257–273.

(18) Penn, R. L. Imperfect Oriented Attachment: Dislocation Generation in Defect-Free Nanocrystals. *Science* **1998**, *281* (5379), 969–971.

(19) Yuwono, V. M.; Burrows, N. D.; Soltis, J. A.; Penn, R. L. Oriented Aggregation: Formation and Transformation of Mesocrystal Intermediates Revealed. *J. Am. Chem. Soc*. **2010**, *132* (7), 2163–2165.

(20) Wu, H.; Yang, Y.; Ou, Y.; Lu, B.; Li, J.; Yuan, W.; Wang, Y.; Zhang, Z. Early Stage Growth of Rutile Titania Mesocrystals. *Cryst. Growth Des.* **2018**, *18* (8), 4209–4214.

(21) Ossorio, M.; Van Driessche, A. E. S.; Pérez, P.; García-Ruiz, J. M. The Gypsum–anhydrite Paradox Revisited. *Chem. Geol.* **2014**, *386*, 16–21.

(22) Van Driessche, A. E. S.; Canals, A.; Ossorio, M.; Reyes, R. C.; García-Ruiz, J. M. Unraveling the Sulfate Sources of (Giant) Gypsum Crystals Using Gypsum Isotope Fractionation Factors. *J. Geol.* **2016**, *124* (2), 235–245.

(23) Wirth, R. Focused Ion Beam (FIB) Combined with SEM and TEM: Advanced Analytical Tools for Studies of Chemical Composition, Microstructure and Crystal Structure in Geomaterials on a Nanometre Scale. *Chem. Geol.* **2009**, *261* (3–4), 217–229.

(24) Wirth, R. Focused Ion Beam (FIB): A Novel Technology for Advanced Application of Micro- and Nanoanalysis in Geosciences and Applied Mineralogy. *Eur. J. Mineral.* **2004**, *16* (6), 863–

876.

(25) Heinemann, S.; Wirth, R.; Gottschalk, M.; Dresen, G. Synthetic [100] Tilt Grain Boundaries in Forsterite: 9.9 to 21.5°. *Phys. Chem. Miner.* **2005**, *32* (4), 229–240.

(26) Rueden, C. T.; Schindelin, J.; Hiner, M. C.; DeZonia, B. E.; Walter, A. E.; Arena, E. T.; Eliceiri, K. W. ImageJ2: ImageJ for the next Generation of Scientific Image Data. *BMC Bioinformatics* **2017**, *18* (1), 529.

(27) Oliphant, T. E. *Guide to NumPy*; 2006; Vol. 1.

(28) Hunter, J. D. Matplotlib: A 2D Graphics Environment. *Comput. Sci. Eng.* **2007**, *9* (3), 99–104.

(29) de la Pena, F.; Fauske, V. T.; Burdet, P.; Prestat, E.; Jokubauskas, P.; Nord, M.; Ostasevicius, T.; MacArthur, K. E.; Sarahan, M.; Johnstone, D. N.; et al. HyperSpy v1.4.1 https://zenodo.org/record/1469364#.XApIly7YquU (accessed Dec 7, 2018).

(30) Malis, T.; Cheng, S. C.; Egerton, R. F. EELS Log-Ratio Technique for Specimen-Thickness Measurement in the TEM. *J. Electron Microsc. Tech.* **1988**, *8* (2), 193–200.

(31) Bellamy, H. D.; Snell, E. H.; Lovelace, J.; Pokross, M.; Borgstahl, G. E. O. The High-Mosaicity Illusion: Revealing the True Physical Characteristics of Macromolecular Crystals. *Acta Crystallogr. Sect. D Biol. Crystallogr.* **2000**, *56* (8), 986–995.

(32) Snell, E. H.; Weisgerber, S.; Helliwell, J. R.; Hölzer, K.; Schroer, K. Improvements in Lysozyme Protein Crystal Perfection through Microgravity Growth. *Acta Crystallogr. D. Biol. Crystallogr.* **1995**, *51* (Pt 6), 1099–1102.

(33) Ferrari, C.; Zanotti, L.; Zappettini, A.; Arumainathan, S. Mosaic GaAs Crystals for Hard X-Ray Astronomy. In *SPIE*; Goto, S., Khounsary, A. M., Morawe, C., Eds.; International Society for

Optics and Photonics, 2008; Vol. 7077, p 70770O–70770O–11.

(34) Lee, M. R. The Petrography, Mineralogy and Origins of Calcium Sulphate within the Cold Bokkeveld CM Carbonaceous Chondrite. *Meteoritics* **1993**, *28* (1), 53–62.

(35) Van Driessche, A. E. S.; Benning, L. G.; Rodriguez-Blanco, J. D.; Ossorio, M.; Bots, P.; García-Ruiz, J. M. The Role and Implications of Bassanite as a Stable Precursor Phase to Gypsum Precipitation. *Science* **2012**, *335* (6077), 69–72.

(36) Tritschler, U.; Van Driessche, A. E. S.; Kempter, A.; Kellermeier, M.; Cölfen, H. Controlling the Selective Formation of Calcium Sulfate Polymorphs at Room Temperature. *Angew. Chemie - Int. Ed.* **2015**, *54* (13), 4083–4086.

(37) Tritschler, U.; Kellermeier, M.; Debus, C.; Kempter, A.; Cölfen, H. A Simple Strategy for the Synthesis of Well-Defined Bassanite Nanorods. *CrystEngComm* **2015**, *17* (20), 3772–3776.

(38) Van Driessche, A. E. S.; Garcia-Ruiz, J. M.; Tsukamoto, K.; Patino-Lopez, L. D.; Satoh, H. Ultraslow Growth Rates of Giant Gypsum Crystals. *Proc. Natl. Acad. Sci.* **2011**, *108* (38), 15721–15726.

(39) Garcia-Guinea, J.; Morales, S.; Delgado, a.; Recio, C.; Calaforra, J. M. M.; García-Guinea, J.; Morales, S.; Delgado, a.; Recio, C.; Calaforra, J. M. M. Formation of Gigantic Gypsum Crystals. *J. Geol. Soc. London.* **2002**, *159* (4), 347–350.

(40) Foshag, W. F. The Selenite Caves of Naica, Mexico. *Am. Mineral.* **1927**, *12* (6), 252–256.

(41) Smeets, P. J. M.; Cho, K. R.; Sommerdijk, N. A. J. M.; De Yoreo, J. J. A Mesocrystal-Like Morphology Formed by Classical Polymer-Mediated Crystal Growth. *Adv. Funct. Mater.* **2017**, *27* (40), 1701658.

(42) Kim, Y.-Y.; Schenk, A. S.; Ihli, J.; Kulak, A. N.; Hetherington, N. B. J.; Tang, C. C.; Schmahl, W. W.; Griesshaber, E.; Hyett, G.; Meldrum, F. C. A Critical Analysis of Calcium Carbonate Mesocrystals. *Nat. Commun.* **2014**, *5* (1), 4341.

(43) Ossorio, M.; Stawski, T.; Rodríguez-Blanco, J.; Sleutel, M.; García-Ruiz, J.; Benning, L.; Van Driessche, A. Physicochemical and Additive Controls on the Multistep Precipitation Pathway of Gypsum. *Minerals* **2017**, *7* (8), 140.

# *Supporting Information*

*to*

# *Particle-mediated nucleation pathways are imprinted in the internal structure of calcium sulfate single crystals*

Tomasz M. Stawski[1]*, Helen M. Freeman[1,2], Alexander E.S. Van Driessche[3]**, Jörn Hövelmann[1], Rogier Besselink[1,3], Richard Wirth[1], and Liane G. Benning[1,4,5]

[1]German Research Centre for Geosciences, GFZ, Interface Geochemistry, Telegrafenberg, 14473, Potsdam, Germany;

[2]School of Chemical and Process Engineering, University of Leeds, Woodhouse Lane, LS2 9JT, Leeds, UK;

[3]Université Grenoble Alpes, Université Savoie Mont Blanc, CNRS, IRD, IFSTTAR, ISTerre, 38000 Grenoble, France ;

[4]Department of Earth Sciences, Free University of Berlin, Malteserstr. 74-100 / Building A, 12249 , Berlin, Germany.

[5]School of Earth and Environment, University of Leeds, Woodhouse Lane, LS2 9JT, Leeds, UK.

Corresponding authors:

* tomasz.stawski@gmail.com; https://www.researchgate.net/profile/Tomasz_Stawski

** alexander.van-driessche@univ-grenoble-alpes.fr

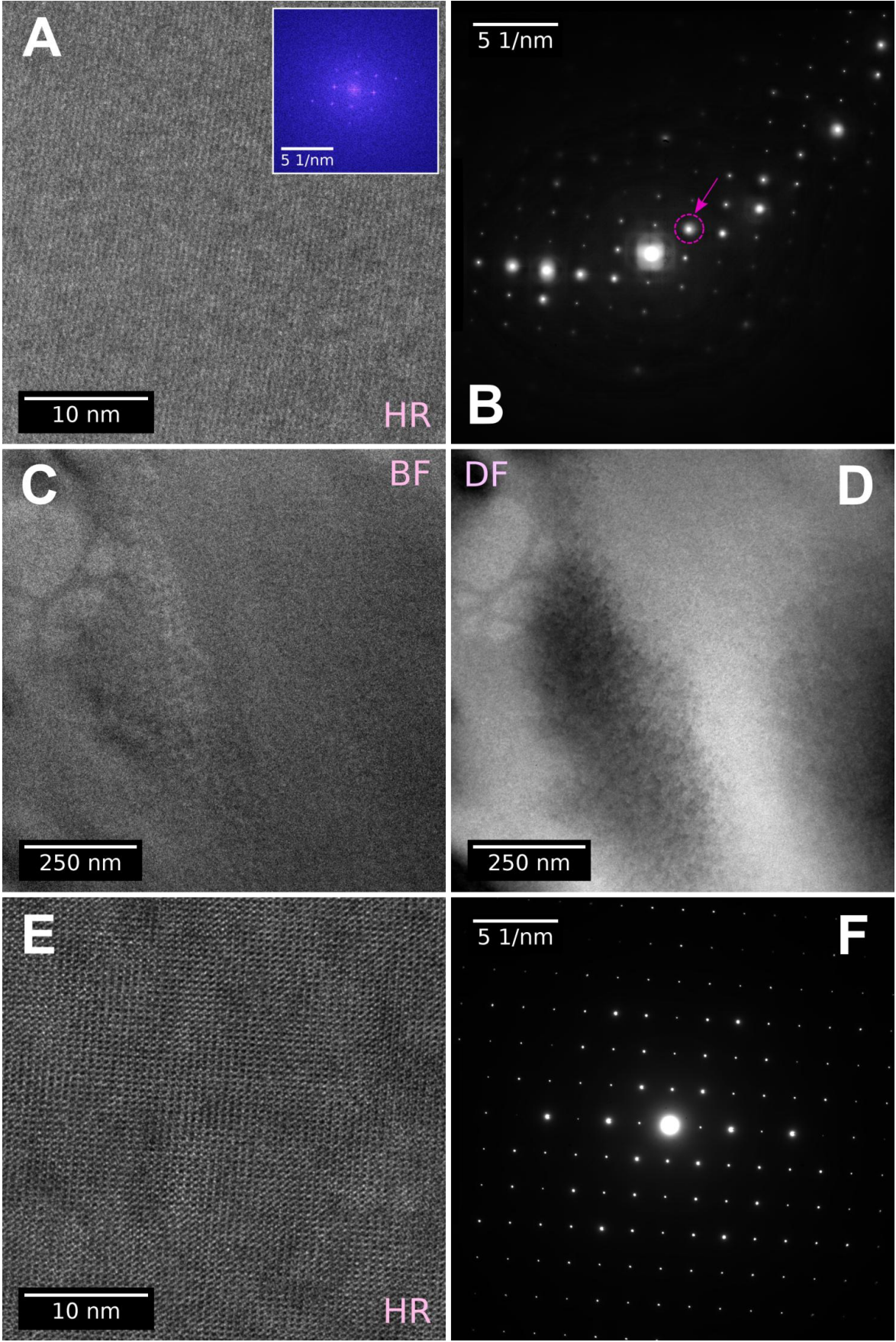

Figure S1. TEM analysis of a FIB thin-foil cut from a natural forsterite (nominally $Mg_2SiO_4$) and generic olivine $((Mg,Fe^{2+})_2SiO_4)$ single crystals. Electron beam current; A) HR image of forsterite showing clear and uniform lattice fringes in the entire field of view; flux: ~8 x $10^5$ $e^-Å^{-2}s^{-1}$, estimated received fluence ~1 x $10^{27}$ $e^-/m^2$; the inset shows the FFT of the HR image and indicates that the lattice fringes originate from a single orientation of a crystal; B) SAED pattern of forsterite with the dashed circle marking the diffracted beam used for dark field imaging; C) low-magnification BF image of forsterite; flux: ~980 $e^-Å^{-2}s^{-1}$, estimated received fluence ~1 x $10^{25}$ $e^-/m^2$ D) DF TEM image corresponding to (C) showing that the field of view is essentially crystallographically uniform, with the differences in contrast originating from the imperfections of the FIB foil, such as its warping; E) HR image of highly-crystalline olivine showing clear and uniform lattice fringes in the entire field of view; flux: ~8 x $10^5$ $e^-Å^{-2}s^{-1}$, estimated received fluence ~1 x $10^{27}$ $e^-/m^2$; F) SAED pattern of olivine demonstrating a high-quality single crystalline nature of the sample.